\documentclass[sigconf,natbib=true,anonymous=false]{acmart}
\AtBeginDocument{%
  }

\copyrightyear{2026}
\acmYear{2026}
\setcopyright{cc}
\setcctype{by}
\acmConference[SIGIR '26]{Proceedings of the 49th International ACM SIGIR Conference on Research and Development in Information Retrieval}{July 20--24, 2026}{Melbourne, VIC, Australia}
\acmBooktitle{Proceedings of the 49th International ACM SIGIR Conference on Research and Development in Information Retrieval (SIGIR '26), July 20--24, 2026, Melbourne, VIC, Australia}
\acmDOI{10.1145/3805712.3809952}
\acmISBN{979-8-4007-2599-9/2026/07}

\usepackage{multirow}
\usepackage{booktabs}
\usepackage{caption}
\usepackage{tcolorbox}
\usepackage{enumitem}
\usepackage{algorithm}
\usepackage{algpseudocode}
\usepackage{array}

\begin{document}

\title{One Pass, Any Order: Position-Invariant Listwise Reranking for LLM-Based Recommendation}

\author{Ethan Bito}
\affiliation{%
  \institution{RMIT University}
  \city{Melbourne}
  \country{Australia}}
\email{s4102812@student.rmit.edu.au}

\author{Yongli Ren}
\affiliation{%
  \institution{RMIT University}
  \city{Melbourne}
  \country{Australia}}
\email{yongli.ren@rmit.edu.au}

\author{Estrid He}
\affiliation{%
  \institution{RMIT University}
  \city{Melbourne}
  \country{Australia}}
\email{estrid.he@rmit.edu.au}

\renewcommand{\shortauthors}{Ethan Bito, Yongli Ren, and Estrid He}


\begin{abstract}
Large language models (LLMs) are increasingly used for recommendation reranking, but their listwise predictions can depend on the order in which candidates are presented. This creates a mismatch between the set-based nature of recommendation and the sequence-based computation of decoder-only LLMs, where permuting an otherwise identical candidate set can change item scores and final rankings. Such order sensitivity makes LLM-based rerankers difficult to rely on, since rankings may reflect prompt serialization rather than user preference. We propose InvariRank, a permutation-invariant listwise reranking framework that addresses this dependence at the architectural level. InvariRank blocks cross-candidate attention with a structured attention mask, and negates position-induced scoring changes through shared positional framing under Rotary Positional Embeddings (RoPE). Combined with a listwise learning-to-rank objective, the model scores all candidates in a single forward pass, avoiding permutation-based invariance training objectives which require multiple permutations of a candidate set. Experiments on recommendation benchmarks show that InvariRank maintains competitive ranking effectiveness while producing stable rankings across candidate permutations. The results suggest that architectural invariance is a practical route to reliable and efficient LLM-based recommendation reranking. The source code is at~\url{https://github.com/ejbito/InvariRank}.
\end{abstract}



\maketitle

\section{Introduction}

Large language models (LLMs) are increasingly used for listwise reranking in recommender systems, as they combine item descriptions, user histories, and task instructions within a single conditional scoring process~\cite{he2020timesan,hou2024large,geng2022recommendation,wei2024llmrec}. In two-stage pipelines, a fixed set of retrieved candidates is passed to the LLM and reranked in one forward pass, making listwise LLM reranking a natural and flexible design choice.

However, this use of LLMs introduces a fundamental mismatch. Recommendation candidates form a set, but decoder-only LLMs process them as a sequence. As a result, rankings can depend not only on user preference and item relevance, but also on the arbitrary order in which candidates are serialized. Even when the user history and candidate set are unchanged, simply permuting the candidate order can produce different scores and rankings~\cite{tang2024found,bito2025evaluatingpositionbiaslarge}. This goes beyond prompt sensitivity, such that the ranker may fail to behave as a well-defined ranking function over a candidate set.

Such order dependence creates a serious reliability problem for LLM-based recommendation. A model that promotes an item in one serialization but demotes it in another is unstable and difficult to evaluate, reproduce, and deploy. Standard ranking metrics can obscure this issue because they measure the quality of a single observed ranking, while ignoring whether the same candidate set would produce a different ranking under an equivalent presentation. For listwise LLM reranking to be reliable, ranking quality and permutation robustness must both be considered. To address this, we enforce permutation invariance directly at the architectural level.

We argue that this sensitivity arises from two architectural sources in decoder-only LLMs. First, cross-candidate attention leakage allows tokens from one candidate to attend to tokens from others, creating order-dependent interference between items. Second, offset drift under Rotary Positional Embeddings (RoPE) changes the relative positional offsets between candidates and the shared user context when candidates are permuted, perturbing attention patterns and item scores~\cite{tang2024found,su2024roformer}. Together, these mechanisms couple predicted relevance to candidate position rather than only to candidate content and user context.

Prior work can reduce this behavior, but generally leaves the source of order dependence intact. Inference-time methods such as permutation ensembling improve stability by aggregating multiple candidate orders, but require repeated model calls~\cite{tang2024found,hou2024largelanguagemodelszeroshot}. Comparison-based prompting avoids some listwise ordering effects by decomposing ranking into smaller decisions, but increases inference cost~\cite{bito2025evaluatingpositionbiaslarge,chao2024makelargelanguagemodel}. Training-time regularization penalizes disagreement across permutations, but the model can still access order information during computation~\cite{chao2024makelargelanguagemodel}.

We instead remove order-dependent channels architecturally. We design the computation so that each candidate is scored from the shared user context and its own content, while preventing other candidates from influencing its representation. We further use a shared positional framing under RoPE so that each candidate is evaluated under the same positional relationship to the user context, regardless of its position in the serialized input~\cite{dong2019unified,su2024roformer}. This makes permutation invariance a property of the model computation rather than an effect of repeated inference or auxiliary loss terms.

We make the following contributions. (1) We identify cross-candidate attention leakage and RoPE-induced offset drift as two concrete sources of position bias in decoder-only LLM reranking. (2) We propose InvariRank, a permutation-invariant, single-pass listwise reranker that enforces set-level invariance by design, while retaining listwise learning through a LambdaRank objective. (3) We show that InvariRank maintains competitive ranking effectiveness while substantially improving permutation robustness across benchmarks, providing a practical path toward reliable LLM-based recommendation reranking.
\section{Related Work}

In recommender systems~\cite{he2019joint,zhu2014effective,zhang2019deep}, LLMs are commonly deployed as rerankers in two-stage pipelines, where a retrieval model first produces a fixed candidate set because of computational and context-length constraints~\cite{luo2024recrankerinstructiontuninglarge,gao2025llm4rerankllmbasedautorerankingframework,zhang-etal-2025-enhancing,sun2023chatgpt}. Among pointwise, pairwise, and listwise formulations, listwise reranking has emerged as a particularly effective paradigm for LLMs, as it enables joint scoring of all candidates and leverages global context~\cite{ma2023zeroshotlistwisedocumentreranking,zhang2023rankwithoutgptbuildinggptindependentlistwise,Zhuang_2024,ma2023zero}.

However, recent work shows that listwise LLM rerankers can produce different rankings for the same candidate set when the input order is changed~\cite{cuconasu2025ragsystemsreallysuffer,hou2024largelanguagemodelszeroshot,lu2022fantastically}. This behavior arises from architectural properties of autoregressive transformers, including causal attention and positional encodings~\cite{cuconasu2025ragsystemsreallysuffer}, which couple predicted relevance to candidate position. As a result, ranking outputs can depend on prompt serialization rather than only on the underlying items and user context.

Prior approaches aim to mitigate this effect but do not remove its source. Inference-time methods such as permutation aggregation and greedy selection improve stability by combining multiple candidate orders, but require repeated model calls~\cite{hou2024largelanguagemodelszeroshot,bito2025evaluatingpositionbiaslarge}. Post-hoc calibration methods adjust scores to counteract position bias after scoring has already been affected~\cite{ma2023largelanguagemodelsstable}. Training-time regularization and position-aware fine-tuning penalize disagreement across permutations, but the model still has access to order information during computation~\cite{zhang2024positionawareparameterefficientfinetuning,chao2024makelargelanguagemodel}. These approaches reduce variability in outputs, but operate outside the model’s core computation and leave the underlying order dependence intact.

A notable training-time method is ALRO~\cite{chao2024makelargelanguagemodel}, which encourages invariance through a loss-based formulation by casting ranking as a generative task (e.g., producing sequences such as \texttt{"A > B > C"}). While effective in explicit-feedback settings, ALRO differs from our setting in two key aspects. It does not operate in a two-stage reranking pipeline over retrieved candidates, and it enforces invariance through training objectives rather than model computation.

In contrast, we enforce permutation invariance directly within the scoring architecture. Our approach removes order-dependent computation in a single forward pass, rather than reducing its effects through aggregation, calibration, or regularization.
\begin{figure}[t]
    \centering
    \caption{Overview of InvariRank. Structured attention blocks cross-candidate interactions, and shared positional framing aligns positional offsets across candidates.}
    \vspace{0.2cm}
    \includegraphics[width=1\linewidth]{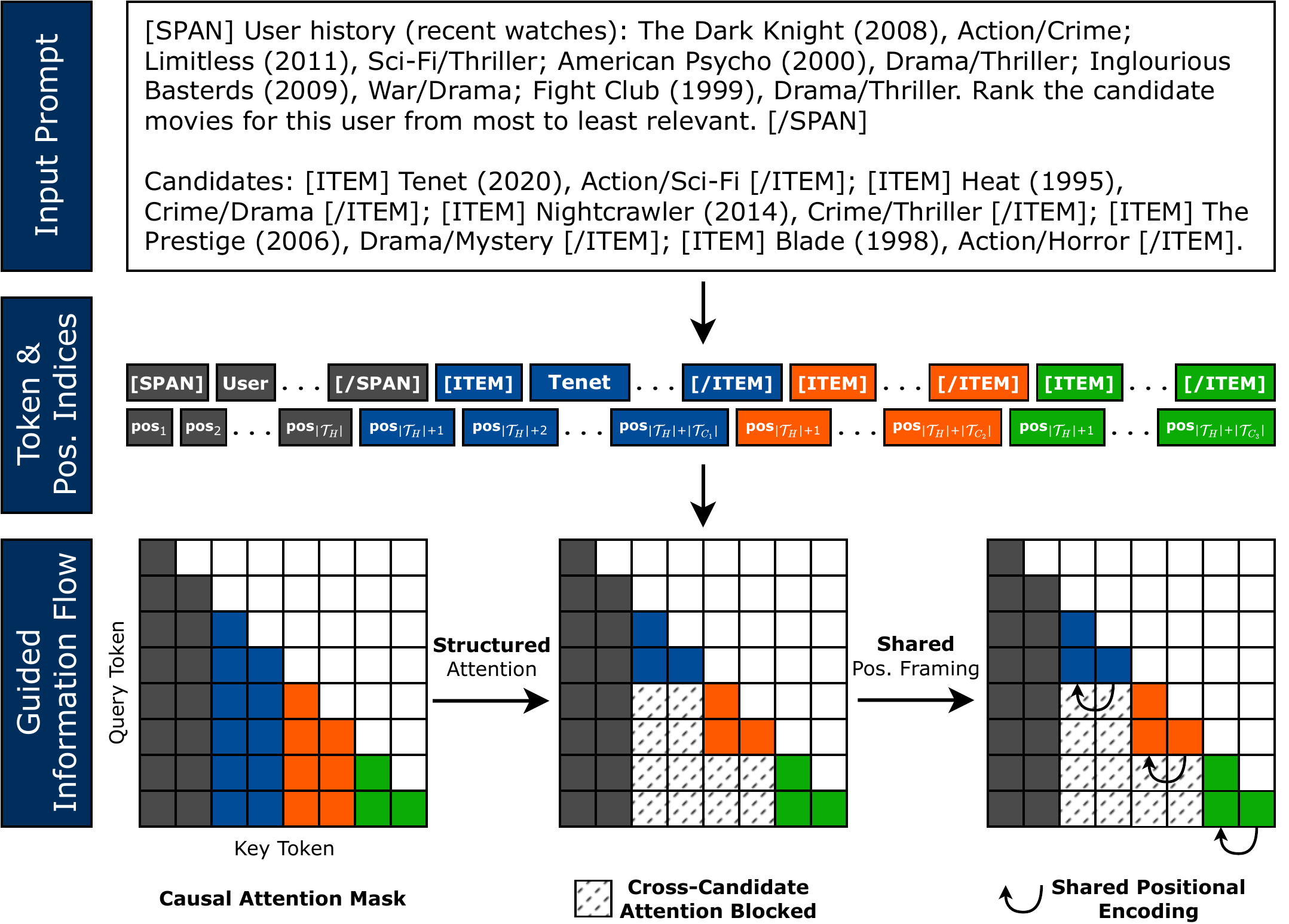}
    \label{fig:top-k-exposure}
    \vspace{-0.5cm}
\end{figure}

\section{Methodology}

We formulate the LLM-based recommendation task as a permutation-equivariant listwise scoring problem, where each candidate receives the same score regardless of input order. Since recommendation operates over sets, scores should not depend on how candidates are serialized. We therefore design the model to produce permutation-invariant scores while retaining listwise supervision via a LambdaRank objective, and introduce InvariRank as a corresponding framework.

\subsection{Permutation-Equivariant Listwise Ranking}

Let $\mathcal{C} = \{c_1, \dots, c_N\}$ denote a candidate set and $\pi$ a permutation over its indices. A reranker should satisfy permutation equivariance
\begin{equation}
s_{\pi(i)}(\mathcal{H}, \pi(\mathcal{C})) = s_i(\mathcal{H}, \mathcal{C}),
\end{equation}
meaning that reordering the input candidates does not change the score assigned to each item. The induced ranking is therefore consistent under permutation.

For each user query, we construct a single input sequence as
\begin{equation}
\small
\begin{array}{c}
q = \texttt{concat}\big(\underbrace{\texttt{[SPAN]}\ \mathcal{I}\ \Vert\ \mathcal{H}\ \texttt{[/SPAN]}}_{\text{shared context}}, \\[4pt]
\phantom{q = \texttt{concat}\big(}\underbrace{\texttt{[ITEM]}\ c_1\ \texttt{[/ITEM]}\ \cdots\ \texttt{[ITEM]}\ c_N\ \texttt{[/ITEM]}}_{\text{candidate set}}\big)
\end{array}
\end{equation}
where $\mathcal{I}$ is the instruction, $\mathcal{H}$ is user history, and $\mathcal{C}$ is the candidate set.

We treat the LLM as a scoring function $f_\theta$. For each candidate $c_i$, we compute a scalar score $s_i$ as the mean token-level log-probability over its \texttt{[ITEM]} span. Candidates are ranked in a single forward pass.

However, standard decoder-only LLMs do not satisfy permutation equivariance.

\subsection{Order Dependence in Decoder-Only LLMs}

Under standard causal attention and positional encoding, candidate scores take the form
\begin{equation}
s_i = f_\theta(\mathcal{H}, c_i, c_{<i}, p_i),
\end{equation}
where $c_{<i}$ denotes preceding candidates and $p_i$ the serialized position. This introduces two sources of order dependence including (i) cross-candidate attention leakage, and (ii) position-dependent encoding under RoPE.

As a result, permuting the input sequence can change scores, violating permutation equivariance.

\subsection{Structured Attention for Candidate Isolation}

To remove cross-candidate interference, we impose a structured segment mask. Let $\mathcal{T}_H$ denote token indices of the shared context and $\mathcal{T}_i$ those of candidate $c_i$. We define
\begin{equation}
M^{\text{seg}}_{t,u}=
\mathbb{I}\Big[(t,u)\in (\mathcal{T}_H\times\mathcal{T}_H)
\ \cup\ \bigcup_{i=1}^{N}(\mathcal{T}_i\times\mathcal{T}_i)
\ \cup\ \bigcup_{i=1}^{N}(\mathcal{T}_i\times\mathcal{T}_H)\Big]
\end{equation}
and apply $M = M^{\text{causal}} \wedge M^{\text{seg}}$.

This restricts each candidate to attend only to the shared context and itself under the causal mask, yielding
\begin{equation}
s_i = g_\theta(\mathcal{H}, c_i),
\end{equation}
removing dependence on other candidates.

\subsection{Shared Positional Framing}

Even with isolated attention, positional encodings can introduce order dependence. To address this, we assign positions using a shared frame.

Tokens in $\mathcal{T}_H$ receive positions $1, \dots, |\mathcal{T}_H|$. Each candidate segment $\mathcal{T}_i$ is assigned
\[
|\mathcal{T}_H|+1, \dots, |\mathcal{T}_H|+|\mathcal{T}_i|
\]
independent of $i$ and the permutation.

For any permutation $\pi$, relative offsets between candidate and context tokens are preserved
\begin{equation}
p_{\pi}(u)-p_{\pi}(t) = p(u)-p(t), \quad \forall t \in \mathcal{T}_H,\ u \in \mathcal{T}_i.
\end{equation}

Thus, each candidate is evaluated under an identical positional relationship to the context, ensuring invariance.

\subsection{Listwise Training Objective}

Although candidate representations are conditionally independent during the forward pass, training remains listwise. We optimize scores jointly using a LambdaRank-style pairwise logistic loss. For each pair $(i,j)$ with $y_i > y_j$, the loss is
\begin{equation}
\ell_{ij}
=
\left|\Delta \mathrm{nDCG}_{ij}\right|
\log\left(1+\exp\left(-\sigma(s_i-s_j)\right)\right).
\end{equation}
The final loss is obtained by averaging over all preference pairs. This weights each pairwise preference by the change in nDCG induced by swapping the two candidates.

This couples candidate scores through the loss, preserving listwise supervision while ensuring that permutation invariance is enforced by the model computation rather than the objective.
\section{Experiment Setup}

\noindent\textbf{Datasets.}
We evaluate on two datasets: MovieLens-32M~\cite{10.1145/2827872} and Amazon Books~\cite{hou2024bridging}, both with timestamped explicit-feedback interactions. For each user, interactions are sorted chronologically and split into $70\%/10\%/20\%$ train/validation/test segments. Each query uses the 20 most recent past interactions as history, and relevance is derived from future interactions to avoid temporal leakage.

\noindent\textbf{Candidate Generation.}
We use LightGCN as a first-stage retriever trained on past interactions with ratings $\geq 4$ as implicit positives. For each query, we construct a reranking list of size $K=25$ from a retrieved candidate pool, ensuring future positives are included when available, and filling remaining slots with retrieved or sampled non-interacted items. All methods rerank the same candidate sets, with non-interacted items assigned relevance $0$.

\begin{table*}[t]
\setlength{\abovecaptionskip}{4pt}
\setlength{\belowcaptionskip}{2pt}
\centering

\footnotesize
\setlength{\tabcolsep}{4pt}
\renewcommand{\arraystretch}{1.1}

\caption{Reranking results on MovieLens-32M \& Amazon Books with list length $K=25$. HR@k and nDCG@k denote Hit Rate and nDCG at $k \in \{5,10\}$, and Kendall’s $\tau$, Spearman’s $\rho$, and top-$k$ agreement (T@5) measure permutation robustness. Higher is better for all metrics.}
\label{tab:main}

\begin{tabular}{c c ccccccc ccccccc}
\hline
\multirow{2}{*}{\textbf{Model}} &
\multirow{2}{*}{\textbf{Method}} &
\multicolumn{7}{c}{\textbf{ML-32M}} &
\multicolumn{7}{c}{\textbf{Books}} \\
& 
& HR@5 & HR@10 & nDCG@5 & nDCG@10 & $\tau$ $\uparrow$ & $\rho$ $\uparrow$ & T@5 $\uparrow$
& HR@5 & HR@10 & nDCG@5 & nDCG@10 & $\tau$ $\uparrow$ & $\rho$ $\uparrow$ & T@5 $\uparrow$ \\
\hline
\noalign{\vskip 2pt}

\multirow{6}{*}{\rotatebox[origin=c]{90}{LLaMA-3B}}
& Zero-shot     & 0.4796 & 0.7323 & 0.1594 & 0.2384 & 0.5364 & 0.6665 & 0.6661 & 0.1426 & 0.3483 & 0.0886 & 0.1540 & 0.5984 & 0.7141 & 0.6828 \\
& Bootstrapping & 0.4986 & 0.7393 & 0.1660 & 0.2448 & 0.6892 & 0.8490 & 0.6596 & 0.1683 & 0.3623 & 0.1051 & 0.1669 & 0.7279 & 0.8685 & 0.6748 \\
& STELLA        & 0.5143 & 0.7936 & 0.1744 & 0.2624 & 0.1176 & 0.1688 & 0.2166 & 0.2309 & 0.4313 & 0.1410 & 0.2048 & 0.0328 & 0.0476 & 0.2084 \\
& SGS           & 0.4993 & 0.7570 & 0.1654 & 0.2456 & 0.7705 & 0.9118 & 0.7476 & 0.1270 & 0.3490 & 0.0808 & 0.1515 & \underline{0.8081} & \underline{0.9106} & \underline{0.8084} \\
\cmidrule(lr){2-16}
& LFT           & \textbf{0.9883} & \textbf{1.0000} & \textbf{0.7679} & \textbf{0.8486} & \underline{0.8861} & \underline{0.9708} & \underline{0.8389} & \textbf{0.4806} & \textbf{0.7096} & \textbf{0.3364} & \textbf{0.4108} & 0.7758 & 0.9020 & 0.7954 \\
& InvariRank    & \underline{0.9700} & \textbf{1.0000} & \underline{0.7260} & \underline{0.8166} & \textbf{0.9883} & \textbf{0.9984} & \textbf{0.9906} & \underline{0.4720} & \underline{0.6833} & \underline{0.3184} & \underline{0.3871} & \textbf{0.9834} & \textbf{0.9977} & \textbf{0.9805} \\[1pt]
\hline
\noalign{\vskip 2pt}

\multirow{6}{*}{\rotatebox[origin=c]{90}{Mistral-7B}}
& Zero-shot     & 0.4480 & 0.7256 & 0.1561 & 0.2332 & 0.4508 & 0.5927 & 0.5406 & 0.1750 & 0.3513 & 0.1127 & 0.1688 & 0.5523 & 0.6944 & 0.6167 \\
& Bootstrapping & 0.4530 & 0.7246 & 0.1572 & 0.2352 & 0.6274 & 0.7964 & 0.5869 & 0.1936 & 0.3616 & 0.1236 & 0.1771 & 0.7018 & 0.8545 & 0.6600 \\
& STELLA        & 0.5180 & 0.8119 & 0.1719 & 0.2624 & 0.0111 & 0.0160 & 0.2053 & 0.2243 & 0.4243 & 0.1428 & 0.2070 & 0.0332 & 0.0475 & 0.2510 \\
& SGS           & 0.5110 & 0.7526 & 0.1786 & 0.2492 & 0.6596 & 0.8287 & 0.5758 & 0.1863 & 0.3666 & 0.1184 & 0.1758 & \underline{0.7432} & \underline{0.8786} & 0.7198 \\
\cmidrule(lr){2-16}
& LFT           & \textbf{0.9936} & \textbf{1.0000} & \textbf{0.7985} & \textbf{0.8698} & \underline{0.8769} & \underline{0.9667} & \underline{0.8419} & \textbf{0.5626} & \textbf{0.7396} & \textbf{0.4049} & \textbf{0.4623} & 0.7407 & 0.8750 & \underline{0.7602} \\
& InvariRank    & \underline{0.9766} & \textbf{1.0000} & \underline{0.7722} & \underline{0.8523} & \textbf{0.9936} & \textbf{0.9992} & \textbf{0.9923} & \underline{0.5216} & \underline{0.6993} & \underline{0.3591} & \underline{0.4166} & \textbf{0.9846} & \textbf{0.9979} & \textbf{0.9843} \\[1pt]
\hline

\end{tabular}
\end{table*}

\noindent\textbf{Implementation.}
We fine-tune LLaMA~3.2~3B-Instruct~\cite{touvron2023llamaopenefficientfoundation} and Mistral~7B-Instruct~\cite{jiang2023mistral7b} using LoRA (rank 16, $\alpha=32$). Each candidate is scored using the mean token-level log-probability over its span. Models are trained with a listwise LambdaRank objective that optimizes pairwise swaps weighted by $\Delta$nDCG. 

We use a maximum sequence length of 4096 and truncate histories to 20 interactions. Training is performed for 500 optimizer steps with batch size 16 (via gradient accumulation), learning rate $5\times10^{-5}$, and AdamW with $5\%$ warmup. Hyperparameters are selected using validation nDCG@10.

For LFT, we use standard causal attention and positional encoding. InvariRank replaces these with structured segment masking and shared positional framing.

\noindent\textbf{Baselines.}
We compare against inference-time, post-hoc, and fine-tuned baselines. \textbf{Zero-shot} applies an LLM reranker without fine-tuning~\cite{Dai_2023,liu2023chatgptgoodrecommenderpreliminary}. \textbf{SGS} constructs rankings greedily with multiple forward passes~\cite{bito2025evaluatingpositionbiaslarge}. \textbf{Bootstrapping} aggregates rankings over permutations~\cite{hou2024largelanguagemodelszeroshot}. \textbf{STELLA} applies post-hoc calibration using a learned position-bias matrix~\cite{ma2023largelanguagemodelsstable}. We adapt STELLA to produce ranked lists in our setting. It is designed to improve ranking effectiveness under position bias, rather than to enforce consistency in the induced preference structure. \textbf{LFT} denotes listwise fine-tuning with the same scoring function and LambdaRank objective as InvariRank but without architectural invariance constraints.

\noindent\textbf{Evaluation.}
We evaluate ranking effectiveness using HR@$k$ and nDCG@$k$ for $k \in \{5,10\}$. Permutation robustness is measured using Kendall’s $\tau$, Spearman’s $\rho$, and top-$k$ agreement across rankings produced from different permutations of the same candidate set. To assess sensitivity, each candidate set is evaluated under multiple permutations and metrics are averaged across permutations.

\begin{figure}[b]
    \centering
    \caption{Top-5 exposure by input position on ML-32M using LLaMA 3.2 3B-Instruct with list length $K=25$.}
    \vspace{0.2cm}
    \includegraphics[width=1\linewidth]{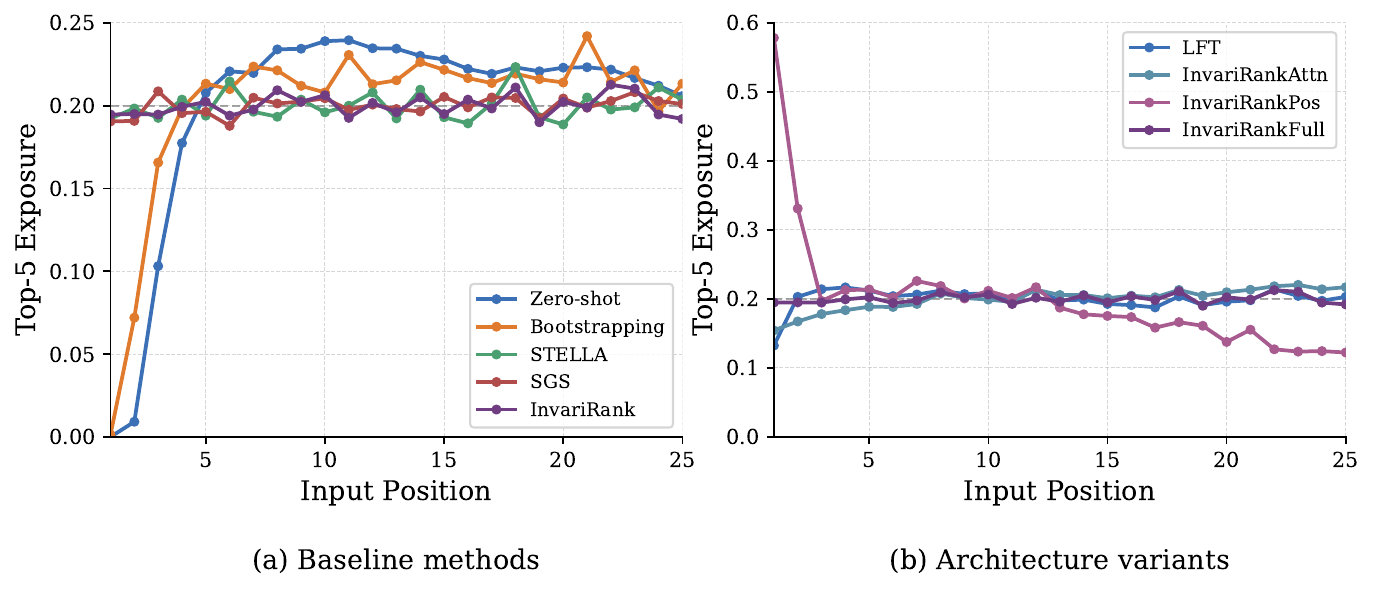}
    \label{fig:top-k-exposure}
    \vspace{-0.5cm}
\end{figure}
\section{Results}

Table~\ref{tab:main} reports recommendation performance and permutation robustness on MovieLens-32M and Amazon Books. We compare InvariRank against prompting-based, post-hoc, and fine-tuned baselines.

\noindent\textbf{Main results.}
Listwise fine-tuning (LFT) achieves the strongest ranking effectiveness across both datasets, substantially improving HR and nDCG over inference-time baselines. However, it remains sensitive to candidate ordering, with lower Kendall’s $\tau$, Spearman’s $\rho$, and top-$k$ agreement. In contrast, InvariRank maintains competitive effectiveness while achieving near-perfect permutation robustness without permutation-based training or ensembling. For example, on MovieLens-32M with LLaMA-3B, InvariRank attains $\tau=0.9883$ and $\rho=0.9984$ compared to $0.8861$ and $0.9708$ for LFT, while retaining comparable nDCG@10 ($0.8166$ vs.\ $0.8486$). Similar trends hold across both backbones and datasets.

\noindent\textbf{Comparison to robustness baselines.}
Inference-time methods such as bootstrapping and SGS improve robustness over zero-shot prompting, but require multiple forward passes and still fall short of InvariRank. STELLA applies post-hoc calibration and is adapted here to produce ranked lists in our setting. Its lower rank correlation reflects its design to improve ranking effectiveness under position bias, rather than to enforce consistency in the induced preference ordering. In contrast, InvariRank enforces invariance within the model computation and achieves higher robustness across all metrics in a single forward pass.

\noindent\textbf{Effectiveness--robustness trade-off.}
We observe a consistent trade-off between ranking effectiveness and permutation robustness. While LFT achieves the highest nDCG, its rankings vary under permutation. InvariRank incurs a small reduction in effectiveness but produces highly stable rankings, with large gains in $\tau$, $\rho$, and top-$k$ agreement. This shows that architectural invariance removes position-induced variability that standard fine-tuning does not address.

\noindent\textbf{Position bias and exposure.}
Figure~\ref{fig:top-k-exposure} shows top-$k$ exposure by input position. Zero-shot and bootstrapping exhibit strong position bias, while LFT reduces but does not eliminate it. SGS and STELLA substantially flatten exposure, showing that inference-time methods can mitigate position bias. InvariRank achieves similarly uniform exposure while also providing near-perfect permutation robustness, indicating that architectural invariance better removes residual order dependence.
\begin{table}[ht]
\centering
\caption{Architectural ablation on ML-32M using LLaMA 3.2 3B Instruct, isolating the contribution of each component.}
\label{tab:ablation}
\vspace{-0.3cm}
\small
\setlength{\tabcolsep}{4pt}
\renewcommand{\arraystretch}{1.1}

\begin{tabular}{c ccccc}
\hline
\textbf{Method} & HR@5 & nDCG@5 & $\tau$ & $\rho$ & T@5 \\
\hline

$\text{InvariRank}_{\text{Pos}}$ & 0.6423 & 0.2465 & 0.5389 & 0.6967 & 0.5429 \\
$\text{InvariRank}_{\text{Attn}}$ & 0.9690 & 0.7298 & 0.9193 & 0.9822 & 0.9097 \\
$\text{InvariRank}_{\text{Full}}$ & 0.9700 & 0.7260 & 0.9883 & 0.9984 & 0.9906 \\

\hline
\end{tabular}
\end{table}
\vspace{-0.5cm}

\section{Ablation}

Table~\ref{tab:ablation} evaluates the contribution of structured attention masking and shared positional framing on MovieLens-32M with LLaMA~3.2~3B. Figure~\ref{fig:top-k-exposure}(b) shows the corresponding exposure patterns.

Using only shared positional framing (InvariRank$_{\text{Pos}}$) leads to poor effectiveness and weak robustness ($\tau=0.5389$, $\rho=0.6967$), with strong position bias in exposure. In contrast, structured attention masking (InvariRank$_{\text{Attn}}$) recovers strong ranking performance and substantially improves robustness ($\tau=0.9193$, $\rho=0.9822$), indicating that preventing cross-candidate interference is the main driver of invariance.

Combining both components (InvariRank$_{\text{Full}}$) achieves the best results, with near-perfect robustness ($\tau=0.9883$, $\rho=0.9984$) and uniform exposure. This shows that attention masking removes most order dependence, while shared positional framing further refines invariance.
\section{Discussion}

Our results highlight a fundamental trade-off between ranking effectiveness and permutation robustness in LLM-based reranking. While listwise fine-tuning (LFT) improves effectiveness, it does not eliminate order dependence, leading to inconsistent rankings under permutation. In contrast, InvariRank enforces permutation invariance at the architectural level, producing stable rankings with near-perfect agreement while maintaining competitive effectiveness. This suggests that architectural constraints, rather than training objectives alone, are necessary for reliable listwise ranking.

A key limitation of InvariRank is that candidate isolation removes cross-candidate interactions, potentially discarding useful comparative signals and limiting fine-grained preference modeling. Additionally, our evaluation focuses on fixed-size candidate sets and two recommendation datasets, leaving open how the approach scales to larger candidate sets or generalizes to other ranking tasks such as information retrieval and retrieval-augmented generation (RAG).
Future work should explore hybrid designs that incorporate controlled cross-candidate interaction while preserving permutation invariance, as well as broader evaluation across domains and candidate set sizes.
\section{Conclusion}
We study position bias in listwise LLM-based recommendation reranking, where candidate permutations can change item scores and rankings. We propose InvariRank, an order-invariant listwise reranker that enforces permutation robustness architecturally through structured segment masking and shared positional framing, without permutation ensembling or auxiliary invariance losses.
Evaluations on MovieLens-32M and Amazon Books with Mistral~7B and LLaMA~3.2~3B show that InvariRank maintains competitive ranking effectiveness while achieving near-perfect permutation robustness, measured by Kendall’s $\tau$, Spearman’s $\rho$, and top-$k$ agreement.
These results show that eliminating order dependence by design is a practical route to stable, single-pass LLM reranking for recommendation.


\bibliographystyle{ACM-Reference-Format}
\bibliography{references}


\end{document}